\documentstyle[12pt]{article}
\catcode`\@=11
\def\equationarray{\stepcounter{equation} \let\@currentlabel=\theequation
 \global\@eqnswtrue \global\@eqcnt\z@ \global\@eqargcnt \z@ \let\\=\@equationcr
 \let\@acol\@arrayacol \let\@classz\@eqnclassz \def\@halignto{to\displaywidth}
 \@equationarray}

\def\@equationarray#1{\setbox\@arstrutbox=\hbox{\vrule 
     height\arraystretch \ht\strutbox
     depth\arraystretch \dp\strutbox
     width\z@}
     \@mkpream{#1} 
     \edef\@preamble{\halign \noexpand\@halignto
      \bgroup \tabskip\z@ \@arstrut \@preamble 
      \hfill\tabskip\@centering &\llap{\@sharp}\tabskip\z@\cr}
     \let\@startpbox\@@startpbox \let\@endpbox\@@endpbox
     \bgroup \let\par\relax
     \let\@sharp## \let\protect\relax \lineskip\z@ \baselineskip\z@
     \tabskip\@centering $$ 
     \@preamble}

\def\@eqnclassz{\ifcase \@lastchclass \@acolampacol \or \@ampacol \or
   \or \or \@addamp \or
   \@acolampacol \or \@firstampfalse \@acol \fi
   \global\advance\@eqargcnt\@ne
   \edef\@preamble{\@preamble
     \ifcase \@chnum 
     \hfil$\relax\displaystyle\tabskip\z@\@sharp$\hfil \or
     $\relax\displaystyle\tabskip\z@\@sharp$\hfil \or 
     \hfil$\relax\displaystyle\tabskip\z@\@sharp$\fi
     \global\advance\@eqcnt\@ne}}

\def\endequationarray{\@equationcr
\egroup\global\advance\c@equation\m@ne 
$$\tabskip\@centering\egroup\global\@ignoretrue}

\def\@equationcr{${\ifnum0=`}\fi\@ifstar{\@xequationcr}{\@xequationcr}}
\def\@xequationcr{
    \@ifnextchar[{\@argequationcr}{\ifnum0=`{\fi}${}\@zequationcr
     \global\@eqnswtrue\global\@eqcnt\z@\cr}}

\def\@argequationcr[#1]{\ifnum0=`{\fi}${}\ifdim #1>\z@
   \@xargequationcr{#1}\else  
   \@yargequationcr{#1}\fi}

\def\@xargequationcr#1{
   \@tempdima #1\advance\@tempdima \dp \@arstrutbox 
   \vrule \@height\z@ \@depth\@tempdima \@width\z@ \@zequationcr
     \global\@eqnswtrue\global\@eqcnt\z@ \cr}

\def\@yargequationcr#1{ \@zequationcr
     \global\@eqnswtrue\global\@eqcnt\z@ \cr\noalign{\vskip #1}}

\def\@zequationcr{\@wargequationcr\if@eqnsw\@eqnnum
    \stepcounter{equation}\global\@eqcnt\z@\fi}

\def\@wargequationcr{\ifnum\@eqcnt<\@eqargcnt
    \@amper \@wargequationcr\fi}

\def\@amper{&}

\newcount\@eqargcnt
\def\eqt#1{Eq.(\ref{#1})}
\def\eqts#1{Eqs.(\ref{#1})}
\def\eqtm#1{(\ref{#1})}
\def\vtheta{\vartheta}
\def\frakd#1#2{{\displaystyle#1\over\displaystyle#2}}

\def\hei{\vphantom{\Bigg|}}

\def\sinp#1{\sin^{#1}\hskip-2pt}

\def\alt{\mathrel{\mathpalette\vereq<}}
\def\agt{\mathrel{\mathpalette\vereq>}}
\def\vereq#1#2{\lower3pt\vbox{\baselineskip1.5pt \lineskip1.5pt
               \ialign{$\mpth#1\hfill##\hfil$\crcr#2\crcr\sim\crcr}}}
\def\mpth{\mathsurround=0pt}
\def\drm{\mathrm{d}}

\def\slash#1{\setbox0=\hbox{$#1$}#1\hskip-\wd0\hbox to\wd0{\hss\sl/\/\hss}}
\def\epsilonslash{\epsilon\kern-.4em/}
\def\kslash{k\kern-.47em/}
\def\Lslash{L\kern-.45em/}
\def\pslash{p\kern-.435em/}
\def\partialslash{\partial\kern-.53em/}
\def\qslash{q\kern-.46em/}
\def\Rslash{R\kern-.6em/}
\def\sslash{s\kern-.44em/}
\def\vslash{v\kern-.47em/}
\def\xislash{\xi\kern-.44em/}
\def\disty{\displaystyle}

\def\MeV{\,\mathrm{M e\hskip-1pt V}}

\newcounter{figs}
\newcounter{refs}
\def\ETHES{E_{\mathrm{th}}} 
\def\ETHNC{E_{\mathrm{th}}} 
\def\RES{R^{\mathrm{ES}}} 
\def\RNC{R^{\mathrm{NC}}} 
\begin{document}
\flushbottom

\pagestyle{empty}
\setcounter{page}{0}
\rightline{\sf DFTT 21/93}
\rightline{\sf hep-ph/9305257}
\rightline{\sf May 1993}
\vspace{1cm}
\begin{center} \LARGE\bf
Towards a Model Independent Treatment
of Future Solar Neutrino Data
\end{center}
\vspace{1cm}
\begin{center} \Large\sf
S.M. Bilenky$^{\mathrm{(a,b,c)}}$
and
C. Giunti$^{\mathrm{(b,c)}\star}$
\end{center}
\medskip
\centerline{\large
(a) Joint Institute of Nuclear Research, Dubna, Russia }
\medskip
\centerline{\large
(b) INFN Torino, Via P. Giuria 1, I--10125 Torino, Italy }
\medskip
\centerline{\large
(c) Dipartimento di Fisica Teorica, Universit\`a di Torino }
\vspace*{0.5in}
\centerline{\Large Abstract }
\bigskip

Possibilities of a model independent treatment
of the data from future real-time solar neutrino experiments
(SNO, Super-Kamiokande and others)
are discussed.
It is shown that in the general case of transitions
of the initial solar $\nu_e$'s
into $\nu_\mu$ and/or $\nu_\tau$
the total flux of initial $^8\mathrm{B}$ neutrinos
and the $\nu_e$ survival probability
can be determined
directly from the experimental data.
Relations between observable quantities are derived,
which,
if confronted with the experimental data,
would allow
to reveal the presence of sterile neutrinos
in the solar neutrino flux on the earth.
Neutrino transitions due to
spin and resonant spin-flavour precession
in the magnetic field of the sun
are shortly discussed.

\vfill

\noindent
{\footnotesize
$^{\scriptstyle\star}$
GIUNTI@TORINO.INFN.IT}

\newpage
\pagestyle{plain}

\section{Introduction}

Solar neutrino experiments
are very important
for investigations of neutrino properties
(masses and mixings,
possible anomalously large magnetic moments, ...)
and for
explorations of the invisible central region of the sun
in which solar energy is produced.
At present,
neutrinos from the sun are detected by three
radiochemical detectors
(Homestake [\ref{DAV90}],
SAGE [\ref{SAGE}]
and
GALLEX [\ref{GAL92A}])
and a real-time water Cherenkov detector
(Kamiokande [\ref{KII}]).
In all four experiments
the detected flux of solar neutrinos
is lower than the flux predicted by the
Standard Solar Model (SSM) [\ref{BAH},\ref{TUR88}].
The most natural explanation
of the possible deficit of solar neutrinos
is provided by neutrino mixing [\ref{PON58}]
which produces resonant MSW [\ref{MSW}]
transitions
of the initial $\nu_e$ into neutrinos of other type
in the sun.
In fact,
all existing solar neutrino data can be described by the MSW mechanism.
Two narrow regions were obtained
(see Ref.[\ref{GAL92B}])
for the parameters
$ \Delta m^2 = m_2^2 - m_1^2 $
and
$ \sinp{2}2\vtheta $
($m_1$, $m_2$ are neutrino masses
and $\vtheta$ is the vacuum mixing angle).
However,
this analysis
is based on the assumption that the SSM
prediction [\ref{BAH},\ref{TUR88}]
of the neutrino fluxes
from different reactions is correct.
For example,
the event rate in radiochemical detectors
is given by
\begin{equation}
N
=
\int_{E_{\mathrm{th}}}
\sigma(E) \,
P_{\nu_e\to\nu_e}(E) \,
\phi_{\nu_e}^{\circ}(E) \,
\drm E
\;,
\label{E01}
\end{equation}
where
$ \sigma(E) $
is the cross section for neutrino capture
with energy threshold
$ E_{\mathrm{th}} $,
$ P_{\nu_e\to\nu_e}(E) $
is the $ \nu_e $ survival probability,
$ \phi_{\nu_e}^{\circ}(E) $
is the initial $ \nu_e $ flux
and $ E $ is the neutrino energy.
At present,
in order to get informations
about the survival probability
$ P_{\nu_e\to\nu_e}(E) $,
which is determined by neutrino properties,
it is necessary to use the
neutrino flux
$ \phi_{\nu_e}^{\circ}(E) $
predicted by the SSM,
including the neutrino flux from $^8\mathrm{B}$ decay,
which
constitutes a very small part
($\sim10^{-4}$ in the SSM)
of the total flux
and
is strongly dependent
from the temperature of the sun core,
from the cross section of different nuclear reactions,
expecially
$ p \, ^7\mathrm{Be} \to \mbox{} ^8\mathrm{B} \, \gamma $,
etc... [\ref{BAH}].

A new generation
of real time solar neutrino experiments
are under preparation and under development:
SNO [\ref{SNO}],
Super-Kamiokande [\ref{SK}],
BOREXINO [\ref{BOREX}],
ICARUS [\ref{ICARUS}],
HELLAZ [\ref{HELLAZ}]
and others.

In the present note
we will consider the general case of transitions
of the initial $\nu_e$'s
into $\nu_\mu$ and/or $\nu_\tau$
due to neutrino mixing and matter effects.
We will show that
future solar neutrino experiments
will allow
to determine the survival probability
$ P_{\nu_e\to\nu_e}(E) $
without assumptions about
the initial neutrino flux
and
to determine
the initial neutrino flux
independently from
what happens to neutrinos
on their way from the sun to the detector.
We will discuss also possible transitions
of solar $\nu_e$'s into sterile left-handed (anti)neutrinos
due to Dirac and Majorana mixing [\ref{BIL}]
and transitions of solar $\nu_e$'s
into sterile right-handed neutrinos (Dirac case)
or active right-handed antineutrinos (Majorana case)
due to spin [\ref{VVO}]
or spin-flavour [\ref{LM88},\ref{AKH88}]
precession
of neutrinos in the magnetic field of the sun.
We will show that these possibilities
can be tested
in future solar neutrino experiments
in a model independent way.

\section{Transitions of solar $\nu_e$'s into $\nu_\mu$ and $\nu_\tau$}

Let us start our discussion with
the SNO experiment,
which will probably be one of the first
new-generation experiments to be realized
(data-taking is scheduled to start in 1995 [\ref{SNO92}]).
SNO is a 1000 tons heavy water Cherenkov detector.
In this experiment solar neutrinos will be detected through
observation of the following three reactions:
\begin{equation}
\nu_e \, d \to e^{-} \, p \, p
\;,
\label{SNO1}
\end{equation}
\begin{equation}
\nu_{x} \, e^{-} \to \nu_{x} \, e^{-}
\;,
\label{SNO2}
\end{equation}
\begin{equation}
\nu_{x} \, d \to \nu_{x} \, p \, n
\;.
\label{SNO3}
\end{equation}
Let us consider first the charged-current (CC) reaction \eqtm{SNO1},
which is expected to yield
about 10 CC events/day
(assuming 1/3 of the SSM flux).
A measurement of the CC event rate
$ n^{\mathrm{CC}}(E) $
as a function of neutrino energy $E$
($ E = T_e + 1.44 \MeV $,
where $T_e$ is the electron kinetic energy)
will allow to determine the
energy spectrum of $\nu_e$
on the earth:
\begin{equation}
\phi_{\nu_e}(E)
=
\frakd{ n^{\mathrm{CC}}(E) }{ \sigma_{\nu_{e}d}^{\mathrm{CC}}(E) }
\;,
\label{E05}
\end{equation}
where
$ \sigma_{\nu_{e}d}^{\mathrm{CC}}(E) $
is the total cross section of process \eqtm{SNO1}.
This cross section
was calculated in Ref.[\ref{CC}].

The electron energy threshold in the SNO experiment
will be $\simeq5\MeV$.
This means that only $^8\mathrm{B}$ and hep neutrinos
can be detected.
According to the SSM,
in the region of high electron energy
($ T_e \agt 13 \MeV $)
the contribution of hep neutrinos to the CC events
is expected to be bigger than
that of $^8\mathrm{B}$ neutrinos.
However,
the contribution of hep neutrinos to the total number of CC events
is expected [\ref{BAH}]
to be much less
than that of $^8\mathrm{B}$ neutrinos ($\ll1\%$).
In the following we will neglect
the contribution of hep neutrinos
to the total counting rates.

The spectrum of initial solar $^8\mathrm{B}$ neutrinos is given by
\begin{equation}
\phi_{\nu_e}^{^8\mathrm{B}}(E)
=
\Phi_{\nu_e}^{^8\mathrm{B}} \, X(E)
\;.
\label{E06}
\end{equation}
The function $X(E)$ is the normalized
$ \disty \left( \int X(E) \, \drm E = 1 \right) $
neutrino spectrum
from the decay
$ ^8\mathrm{B} \to \mbox{} ^8\mathrm{Be} + e^{+} + \nu_e $,
which is determined by the phase space factor
(corrections due to forbidden transitions
where calculated in Ref.[\ref{BAH86}]).
The factor
$\Phi_{\nu_e}^{^8\mathrm{B}}$
in \eqt{E06}
is the total flux of initial $^8\mathrm{B}$ solar $\nu_e$'s.
The value of
$\Phi_{\nu_e}^{^8\mathrm{B}}$
depends on the temperature of the sun's core,
the value of the cross sections
of the nuclear reactions of the $pp$ cycle,
etc...
The distortions of the neutrino spectra are
negligibly small under solar conditions [\ref{BAH91}].

A measurement of the energy spectrum
of detected $\nu_e$'s
will allow to determine
the $\nu_e$ survival probability
$ P_{\nu_e\to\nu_e}(E) $
up to a constant.
In fact,
using \eqt{E06},
we have
\begin{equation}
P_{\nu_e\to\nu_e}(E)
=
\frakd{ 1 }{ \Phi_{\nu_e}^{^8\mathrm{B}} } \,
\frakd{ \phi_{\nu_e}(E) }{ X(E) }
\;.
\label{E07}
\end{equation}
A dependence of
$ \disty \frakd{ \phi_{\nu_e}(E) }{ X(E) } $
on the energy $E$
would be an evidence in favor of
resonant transitions of neutrinos in matter
(or just-so vacuum neutrino oscillations).
A detailed investigation of the dependence
of the survival probability on $E$
will allow to distinguish
different mechanisms of neutrino transitions
(for a review of possible mechanisms see,
for example,
Ref.[\ref{PAL92}]).

It is clear that the detection of solar neutrinos
only through the observation of the CC process \eqtm{SNO1}
will not allow us to know
if there are $\nu_\mu$ and/or $\nu_\tau$
in the solar neutrino flux on the earth.
In order to get this information
it is necessary to detect solar neutrinos
also through the observation of neutral current (NC) induced
processes [\ref{NC}].

From the detection of solar neutrinos
through the NC process \eqtm{SNO3}
it is possible to determine
the total flux of initial $^8\mathrm{B}$ neutrinos
$\Phi_{\nu_e}^{^8\mathrm{B}}$.
According to the SSM,
the expected rate of NC events
in the SNO detector
is about 10/day.
The detection threshold
is equal to the deuterium binding energy $\simeq2.2\MeV$.
The rate of NC events is given by
\begin{equation}
N^{\mathrm{NC}}
=
\int_{\ETHNC}
\sigma_{\nu d}^{\mathrm{NC}}(E) \,
\sum_{\ell=e,\mu,\tau} \phi_{\nu_\ell}(E) \,
\drm E
\;,
\label{E14}
\end{equation}
where
$ \sigma_{\nu d}^{\mathrm{NC}}(E) $
is the cross section for the process
$ \nu_{x} \, d \to \nu_{x} \, n \, p $
(for the calculation of
$ \sigma_{\nu d}^{\mathrm{NC}}(E) $
see Ref.[\ref{BAH88}] and references therein).

We assume here that
solar $\nu_e$'s
can transform only into $\nu_\mu$ and $\nu_\tau$.
In this case, we have
\begin{equation}
\sum_{\ell=e,\mu,\tau} P_{\nu_e\to\nu_\ell}(E) = 1
\;,
\label{E29}
\end{equation}
where
$ P_{\nu_e\to\nu_\ell}(E) $
is the probability of the transition
$\nu_e\to\nu_\ell$
($\ell=e,\mu,\tau$).
From \eqt{E29},
it follows that,
independently from the mechanism of possible transitions
of the initial $\nu_e$'s into $\nu_\mu$ and/or $\nu_\tau$,
we have
\begin{equation}
\sum_{\ell=e,\mu,\tau}
\phi_{\nu_\ell}(E)
=
\phi_{\nu_e}^{^8\mathrm{B}}(E)
\;.
\label{E12}
\end{equation}
With the help of
\eqts{E06}, \eqtm{E14} and \eqtm{E12},
for the total flux of initial
$^8\mathrm{B}$ $\nu_e$'s
we obtain
the following relation
\begin{equation}
\Phi_{\nu_e}^{^8\mathrm{B}}
=
\frakd{ N^{\mathrm{NC}} }{ X^{\mathrm{NC}}_{\nu d} }
\;,
\label{E22}
\end{equation}
where
\begin{equation}
X^{\mathrm{NC}}_{\nu d}
=
\int_{\ETHNC}
\sigma_{\nu d}^{\mathrm{NC}}(E) \,
X(E) \,
\drm E
\;.
\label{E20}
\end{equation}
Thus,
in the SNO experiment
the total flux of $^8\mathrm{B}$ neutrinos
will be {\bf measured}.
A comparison of the measured flux
with that predicted by the SSM
will be an important test
of the model.

The detection of solar neutrinos
through observation
of CC and NC events
will allow to determine the survival probability
$ P_{\nu_e\to\nu_e}(E) $
directly from the experimental data:
\begin{equation}
P_{\nu_e\to\nu_e}(E)
=
\beta(E) \,
\frakd{ n^{\mathrm{CC}}(E) }{ N^{\mathrm{NC}} }
\;,
\label{E30}
\end{equation}
where
\begin{equation}
\beta(E)
=
\frakd{
X^{\mathrm{NC}}_{\nu d}
}{
\sigma^{\mathrm{CC}}_{\nu_e d}(E) \,
X(E)
}
\label{E31}
\end{equation}
is a known quantity.

The sensitivity of the NC events
to all types of active neutrinos
is the same
($\nu_e$-$\nu_\mu$-$\nu_\tau$ universality).
For the relative contribution of $\nu_\mu$ and/or $\nu_\tau$
to NC events
we have
\begin{equation}
\RNC
=
1
-
\frakd{
\int_{\ETHNC}
\sigma_{\nu d}^{\mathrm{NC}}(E) \,
P_{\nu_e\to\nu_e}(E) \,
X(E) \,
\drm E
}{
X^{\mathrm{NC}}_{\nu d}
}
\;,
\label{E15}
\end{equation}
where $\phi_{\nu_e}(E)$
is the $\nu_e$ flux
obtained from the observation of CC events.
Let us stress that the ratio $\RNC$
does not depend on the total flux
of initial $^8\mathrm{B}$ $\nu_e$'s.

If the ratio
$ \RNC > 0 $,
it would mean that
$\nu_\mu$ and/or $\nu_\tau$
are present
in the flux of solar neutrinos on the earth.
Notice that
this remains valid
if the $\nu_e$ produced in the sun
transform
not only into $\nu_\mu$ and $\nu_\tau$
but also into sterile states.

Let us consider now the process \eqtm{SNO2}
of elastic scattering (ES)
of neutrinos on electrons,
which is due to CC and NC interactions.
The rate of ES events expected in the SNO detector
will be about 1 event/day
(assuming 1/3 of the SSM flux)
for an electron energy threshold $T_e^{\mathrm{th}}=5\MeV$.
The total rate of ES events is given by
\begin{equation}
N^{\mathrm{ES}}
=
\int_{\ETHES}
\sigma_{\nu_e e}(E) \,
\phi_{\nu_e}(E) \,
\drm E
+
\int_{\ETHES}
\sigma_{\nu_\mu e}(E)
\sum_{\ell=\mu,\tau} \phi_{\nu_\ell}(E) \,
\drm E
\;,
\label{E27}
\end{equation}
where
$ \sigma_{\nu_\ell e}(E) $
is the cross section of the process
$ \nu_\ell \, e \to \nu_\ell \, e $
(for $\ell=e,\mu$)
integrated over the kinetic energy $T_e$ of the recoil electron
in the interval
$ \disty
T_e^{\mathrm{th}} \le T_e \le E / \left( 1 + \frakd{ m_e }{ 2 E } \right)
$
and
$ \disty
E_{\mathrm{th}}
=
\frakd{1}{2} T_e^{\mathrm{th}}
+
\frakd{1}{2}
\sqrt{ T_e^{\mathrm{th}} \left( T_e^{\mathrm{th}} + 2 m_e \right) }
$.
Since
\begin{equation}
\frakd{1}{7} \, \sigma_{\nu_e e}(E)
\alt
\sigma_{\nu_\mu e}(E)
\alt
\frakd{1}{6} \, \sigma_{\nu_e e}(E)
\;,
\label{E32}
\end{equation}
(depending on $T_e^{\mathrm{th}}$)
the contribution of $\nu_\mu$ and/or $\nu_\tau$
to the total rate of ES events
is considerably smaller than the contribution of $\nu_e$.
However,
from the MSW analysis of existing data
it follows that
the relative contribution of the second term in \eqt{E27}
could be as big as 40\%
(see Fig.\ref{F2}).
This means that,
in spite of the suppression due to the cross section ratio \eqtm{E32},
it could be possible to reveal
the presence of $\nu_\mu$ and/or $\nu_\tau$
in the flux of solar neutrinos on the earth
from the observation of ES events.
The relative contribution
of $\nu_\mu$ and/or $\nu_\tau$
to the total rate of ES events
is given by
\begin{equation}
\RES
=
1
-
\frakd{
\int_{\ETHES}
\sigma_{\nu_e e}(E) \,
P_{\nu_e\to\nu_e}(E) \,
X(E) \,
\drm E
}{
\int_{\ETHES}
\sigma_{\nu_e e}(E) \,
P_{\nu_e\to\nu_e}(E) \,
X(E) \,
\drm E
+
\int_{\ETHES}
\sigma_{\nu_\mu e}(E)
\left( 1 - P_{\nu_e\to\nu_e}(E) \right)
X(E) \,
\drm E
}
\;.
\label{E33}
\end{equation}

A measurement of the value of $N^{\mathrm{ES}}$
will also allow to determine the flux of
initial $^8\mathrm{B}$ $\nu_e$'s.
In fact,
from \eqts{E06} and \eqtm{E27},
it is easy to obtain the following expression
for the total flux of initial
$^8\mathrm{B}$ $\nu_e$'s:
\begin{equation}
\Phi_{\nu_e}^{^8\mathrm{B}}
=
\frakd{ \Sigma^{\mathrm{ES}} }{ X^{\mathrm{ES}}_{\nu_\mu e} }
\;,
\label{E13}
\end{equation}
where
\begin{equationarray}{l} \displaystyle \hei
\Sigma^{\mathrm{ES}}
=
N^{\mathrm{ES}}
-
\int_{\ETHES}
\left( \sigma_{\nu_e e}(E) - \sigma_{\nu_\mu e}(E) \right)
\phi_{\nu_e}(E) \,
\drm E
\;,
\label{E09}
\\ \displaystyle \hei
X^{\mathrm{ES}}_{\nu_\mu e}
=
\int_{\ETHES}
\sigma_{\nu_\mu e}(E) \,
X(E) \,
\drm E
\;.
\label{E18}
\end{equationarray}
So the total flux of initial
$^8\mathrm{B}$ $\nu_e$'s
can be determined directly from the experimental data
either through the observation of NC events (\eqt{E22})
or through the observation of ES and CC events (\eqt{E13}).
Let us notice that the quantities
$ N^{\mathrm{NC}} $ and $ \Sigma^{\mathrm{ES}} $,
which can be determined through the observation
of the processes \eqtm{SNO1}--\eqtm{SNO3},
are not independent.
In fact,
independently on the flux of initial
$^8\mathrm{B}$ $\nu_e$'s
we have the following relation:
\begin{equation}
N^{\mathrm{NC}} = 
\frakd{ X^{\mathrm{NC}}_{\nu d} }{ X^{\mathrm{ES}}_{\nu_\mu e} }
\Sigma^{\mathrm{ES}}
\;.
\label{E35}
\end{equation}
As we will show below,
this relation
can be violated only
if sterile neutrinos
or active antineutrinos
are present in the solar neutrino flux on the earth.

Up to now we considered the case in which
only the total rate of ES events will be measured.
Let us assume now that from the experimental data
it will be possible also to obtain
the ES event rate $n^{\mathrm{ES}}(E)$
as a function of the neutrino energy $E$.
The statistics of ES events in the SNO experiment
possibly will not allow to determine $n^{\mathrm{ES}}(E)$
with sufficient accuracy.
However,
a large number of ES events
will be observed in
Super-Kamiokande [\ref{SK}],
ICARUS [\ref{ICARUS}]
and other future solar neutrino experiments.

A measurement of
$n^{\mathrm{CC}}(E)$
and
$n^{\mathrm{ES}}(E)$
will allow to determine
the spectrum of $\nu_\mu$ and/or $\nu_\tau$
directly from the experimental data.
In fact, we have
\begin{equation}
\sum_{\ell=\mu,\tau} \phi_{\nu_\ell}(E)
=
\frakd{ 1 }{ \sigma_{\nu_\mu e}(E) }
\left[
n^{\mathrm{ES}}(E)
-
\sigma_{\nu_e e}(E) \,
\phi_{\nu_e}(E)
\right]
\;.
\label{E25}
\end{equation}
It is clear that
a determination of the fluxes
$ \phi_{\nu_e}(E) $
and
$ \disty \sum_{\ell=\mu,\tau} \phi_{\nu_\ell}(E) $
through the observation of CC and ES events
will allow to predict the rate of NC events
(see \eqt{E14}).
Let us stress that this prediction does not depend
on the possible presence of sterile neutrinos
in the solar neutrino flux on the earth.

\section{Sterile (anti)neutrinos}

In the general case of neutrino mixing,
(Dirac and Majorana mass term [\ref{BIL}])
besides the active neutrinos
$\nu_e$, $\nu_\mu$ and $\nu_\tau$,
sterile neutrinos
$\nu^{\mathrm{S}}_\ell$,
which do not take part in the standard weak interactions,
could exist.
In this case,
the neutrinos with definite mass are Majorana particles
and the number of massive neutrinos
is larger than the number of lepton flavours.
A unique feature of the SNO and other future solar neutrino experiments
is that directly from the data of these experiments
it will be possible to get informations
not only about the presence of $\nu_\mu$ and/or $\nu_\tau$
in the solar neutrino flux,
but also about the presence of sterile (anti)neutrinos.
In fact,
from the unitarity of the mixing matrix,
in the general case of Dirac and Majorana mixing we have
\begin{equation}
\sum_{\ell=e,\mu,\tau} P_{\nu_e\to\nu_\ell}(E)
+
\sum_{\ell} P_{\nu_e\to\nu^{\mathrm{S}}_\ell}(E)
=
1
\;,
\label{E23}
\end{equation}
where
$ \disty \sum_{\ell} P_{\nu_e\to\nu^{\mathrm{S}}_\ell}(E) $
is the total probability of transitions
of $\nu_e$ into all possible sterile states.

From \eqt{E23}
we have
\begin{equation}
N^{\mathrm{NC}}
=
\Phi_{\nu_e}^{^8\mathrm{B}}
X^{\mathrm{NC}}_{\nu d}
-
\int_{\ETHNC}
\sigma_{\nu d}^{\mathrm{NC}}(E) \,
\phi_{\nu^{\mathrm{S}}}(E) \,
\drm E
\;,
\label{E19}
\end{equation}
where
$ \disty
\phi_{\nu^{\mathrm{S}}}(E) =
\sum_{\ell} P_{\nu_e\to\nu^{\mathrm{S}}_\ell}(E) \,
\phi_{\nu_e}^{^8\mathrm{B}}(E)
$
is the flux of sterile neutrinos on the earth
and
$ X^{\mathrm{NC}}_{\nu d} $
is given by \eqt{E20}.
It is clear that without a definite assumption
about
$ \Phi_{\nu_e}^{^8\mathrm{B}} $
it is impossible to make any conclusion
about the existence of sterile neutrinos.
However, we can obtain another relation
for the total flux
$ \Phi_{\nu_e}^{^8\mathrm{B}} $:
\begin{equation}
\Sigma^{\mathrm{ES}}
=
\Phi_{\nu_e}^{^8\mathrm{B}}
X^{\mathrm{ES}}_{\nu_\mu e}
-
\int_{\ETHES}
\sigma_{\nu_\mu e}(E) \,
\phi_{\nu^{\mathrm{S}}}(E) \,
\drm E
\;,
\label{E17}
\end{equation}
where
$ \Sigma^{\mathrm{ES}} $
and
$ X^{\mathrm{ES}}_{\nu_\mu e} $
are given in \eqtm{E09} and \eqtm{E18},
respectively.
Excluding
$ \Phi_{\nu_e}^{^8\mathrm{B}} $
from \eqts{E19} and \eqtm{E17}
we obtain
\begin{equation}
\frakd{ N^{\mathrm{NC}} }{ X^{\mathrm{NC}}_{\nu d} }
-
\frakd{ \Sigma^{\mathrm{ES}} }{ X^{\mathrm{ES}}_{\nu_\mu e} }
=
\frakd{
\int_{\ETHES}
\sigma_{\nu_\mu e}(E) \,
\phi_{\nu^{\mathrm{S}}}(E) \,
\drm E
}{
X^{\mathrm{ES}}_{\nu_\mu e}
}
-
\frakd{
\int_{\ETHNC}
\sigma_{\nu d}^{\mathrm{NC}}(E) \,
\phi_{\nu^{\mathrm{S}}}(E) \,
\drm E
}{
X^{\mathrm{NC}}_{\nu d}
}
\;.
\label{E21}
\end{equation}
The left hand side of \eqt{E21}
contains only quantities
which can be measured through the observation
of CC, NC and ES reactions.
If it is found to be different from zero,~\footnote
{
According to our model calculations
the two terms in the right-hand side
of \eqt{E21}
can cancel each other.
This cancelation depends
on the value of the threshold energy
of the recoil electron in the ES process.
From our calculations
it follows that higher threshold energies are preferable
for the test of the presence
of sterile neutrinos in the solar neutrino flux
with the help of \eqt{E21}.
}
it would follow that there are sterile particles
in the flux of solar neutrinos on the earth.

If the rate
$n^{\mathrm{ES}}(E)$
will be measured,
then the hypothesis of existence of sterile neutrinos
could be tested from the investigation
of only
the CC process \eqtm{SNO1} and
the ES process \eqtm{SNO2}.
In fact,
with the help of \eqt{E23},
we obtain the following relation
\begin{equation}
\frakd{ 1 }{ X(E) }
\left[
\phi_{\nu_e}(E)
+
\sum_{\ell=\mu,\tau} \phi_{\nu_\ell}(E)
\right]
=
\Phi_{\nu_e}^{^8\mathrm{B}}
\left[
1
-
\sum_{\ell} P_{\nu_e\to\nu^{\mathrm{S}}_\ell}(E)
\right]
\;.
\label{E26}
\end{equation}
If it will occur that
the left-hand side of this equation,
which contains only measurable quantities,
depends on energy,
then it would mean that there are sterile neutrinos
in the flux of solar neutrinos on the earth.

To conclude the discussion of tests of the possible existence
of sterile neutrinos,
let us make the following remark:
Assume that
\begin{equation}
\RES = 0
\qquad \hbox{and} \qquad
\RNC = 0
\label{E24}
\end{equation}
but
$ \disty \frakd{ \phi_{\nu_e}(E) }{ X(E) } $
depends on the neutrino energy.
This would mean that in the flux of solar neutrinos on the earth
there are no $\nu_\mu$ and $\nu_\tau$
but the spectrum of $\nu_e$ is distorted
with respect to the initial spectrum
of $^8\mathrm{B}$ neutrinos.
This is possible only in the case
of transitions of $\nu_e$ into sterile neutrinos.

\section{Spin and spin-flavour precession}

Up to now we have considered only
effects due to neutrino mixing.
In the recent years many papers
have discussed possible effects of large neutrino
magnetic moments
($ \simeq 10^{-11} $--$ 10^{-10} \mu_{\mathrm{B}} $,
where $ \mu_{\mathrm{B}} $ is the Bohr magneton)
for solar neutrinos [\ref{VVO},\ref{LM88},\ref{AKH88}].
The future real time solar neutrino experiments
undoubtedly will allow to test this hypothesis
in detail.
We will limit ourselves only to a few remarks.

In the general case of Dirac diagonal and transition magnetic moments
and neutrino mixing,
the flux of solar neutrinos on the earth
contains
left-handed active neutrinos $\nu_{\ell}$ (with $\ell=e,\ldots$)
and
right-handed sterile neutrinos $\nu_{\ell R}$ (with $\ell=e,\ldots$)
produced by
spin and resonant spin-flavour precessions
in the magnetic field of the sun.
In this case,
as in the case of oscillations of
$\nu_e$ into left-handed sterile (anti)neutrinos considered above,
the relations \eqtm{E21} will be satisfied
(where $\phi_{\nu^{\mathrm{S}}}(E)$ is the total flux of
right-handed sterile neutrinos).
So if the left hand side of \eqt{E21},
in which only measurable quantities enter,
is different from zero,
this will be an evidence that in the flux of solar neutrinos
on the earth there are left-handed sterile (anti)neutrinos
(Dirac and Majorana mixing)
or right-handed sterile neutrinos
(Dirac magnetic moments).
It is possible to distinguish these two cases only if
the CC, NC and ES rates depend on time
(which will be an evidence in favour of magnetic moments).

In the case of Majorana
transition magnetic moments,
the flux of solar neutrinos on the earth
contains
left-handed neutrinos,
right-handed antineutrinos
$\bar\nu_\mu$ and $\bar\nu_\tau$
and also possibly
$\bar\nu_e$
[\ref{LM88},\ref{AKH93}].
It is possible to show that a detection
of solar neutrinos through the simultaneous
observation of CC, NC and ES events
will allow to reveal the existence of
antineutrinos in the flux of solar neutrinos on the earth.
In fact,
using only the fact that the total flux of
neutrinos and antineutrinos on the earth must be equal to
the initial flux of $^8\mathrm{B}$ neutrinos,
we obtain the following relation:
\begin{equation}
\begin{array}{rcl} \displaystyle \hei
\frakd{ N^{\mathrm{NC}} }{ X^{\mathrm{NC}}_{\nu d} }
-
\frakd{ \Sigma^{\mathrm{ES}} }{ X^{\mathrm{ES}}_{\nu_\mu e} }
& = & \displaystyle \hei
\frakd{ 1 }{ X^{\mathrm{NC}}_{\nu d} }
\int_{\ETHNC}
\sigma_{\bar\nu d}^{\mathrm{NC}}(E)
\left(
1
-
r_{\nu d}(E)
\right)
\sum_{\ell=e,\mu,\tau} \phi_{\bar\nu_\ell}(E) \,
\drm E
\\ \displaystyle \hei
& & \displaystyle \hei
\mbox{}
-
\frakd{ 1 }{ X^{\mathrm{ES}}_{\nu_\mu e} }
\left[
\int_{\ETHES}
\sigma_{\bar\nu_e e}(E)
\left(
1
-
r_{\bar\nu_e e}(E)
\right)
\phi_{\bar\nu_e}(E) \,
\drm E
\phantom{ \sum_{\ell=\mu,\tau} }
\right.
\\ \displaystyle \hei
& & \displaystyle \hei
\phantom{ X^{\mathrm{NC}}_{\nu d} }
\left.
+
\int_{\ETHES}
\sigma_{\bar\nu_\mu e}(E)
\left(
1
-
r_{\bar\nu_\mu e}(E)
\right)
\sum_{\ell=\mu,\tau} \phi_{\bar\nu_\ell}(E) \,
\drm E
\right]
\;,
\end{array}
\label{E34}
\end{equation}
where
$ X^{\mathrm{NC}}_{\nu d} $,
$ \Sigma^{\mathrm{ES}} $,
and
$ X^{\mathrm{ES}}_{\nu_\mu e} $
are given in \eqts{E20}, \eqtm{E09} and \eqtm{E18},
respectively,
$
r_{\nu d}(E)
=
\sigma_{\nu d}^{\mathrm{NC}}(E) / \sigma_{\bar\nu d}^{\mathrm{NC}}(E)
$,
$
r_{\bar\nu_e e}(E)
=
\sigma_{\nu_\mu e}(E) / \sigma_{\bar\nu_e e}(E)
$,
$
r_{\bar\nu_\mu e}(E)
=
\sigma_{\nu_\mu e}(E) / \sigma_{\bar\nu_\mu e}(E)
$
and
$ \phi_{\bar\nu_\ell} $
are the fluxes of antineutrinos
$\bar\nu_\ell$,
with $\ell=e,\mu,\tau$.
In SNO and other future solar neutrino experiments
it is planned to search for $\bar\nu_e$
from the sun
through the observation
of the reaction
$ \bar\nu_e \, p \to e^{+} \, n $.
Therefore,
with the help of \eqt{E34},
the left hand side of which contains
only quantities
measurable through the observation
of CC, NC and ES reactions,
it will be possible to check
whether $\bar\nu_\mu$ and/or $\bar\nu_\tau$
are present
in the solar neutrino flux on the earth
(investigations of the time dependence
of $N^{\mathrm{NC}}$ and $\Sigma^{\mathrm{ES}}$
are also necessary).
In the energy region discussed here
$ r_{\bar\nu d}(E) $
differs from 1 by only 2--3\% [\ref{BAH88}].
This means that the possible contribution
of the first term in the right-hand side of \eqt{E34}
is strongly suppressed.
We have also
$ -1 \alt 1 - r_{\bar\nu_e e}(E) \alt 0.4 $
and
$ 1 - r_{\bar\nu_\mu e}(E) \simeq - 0.3 $.

Notice that the relation \eqtm{E22}
that connects the total flux
of initial $^8\mathrm{B}$ neutrinos
with the NC event rate $N^{\mathrm{NC}}$
is valid also in the case
of transitions of $\nu_e$
into active neutrinos and antineutrinos.
In order to see this it is necessary to take into account
that
$ \sigma_{\nu d}^{\mathrm{NC}}(E) \simeq
\sigma_{\bar\nu d}^{\mathrm{NC}}(E) $.

\section{Conclusions}

The experiments on the detection of solar neutrinos
have exceptional significance
for the investigation of neutrino properties
(masses, mixings, magnetic moments, ...)
and for the investigations of the sun.
The analysis of the existing solar neutrino data
is based on the assumption
that the standard solar model correctly predicts
the neutrino fluxes from the different reactions
of the $pp$ and CNO cycles.
In the present note
it was shown
that future real-time solar neutrino experiments
(SNO, Super-Kamiokande and others)
will allow to determine directly from the experimental data
the $\nu_e$ survival probability
as well as the flux of initial $^8\mathrm{B}$ neutrinos.
Different tests of the presence in the solar neutrino flux
on the earth of sterile neutrinos
(due to Dirac and Majorana mixing
or spin precession of neutrinos with Dirac magnetic moments)
or active antineutrinos
(due to resonance spin-flavour precession
of neutrinos with Majorana magnetic moments)
are proposed.

The realization of the program
of determination
of the initial $^8\mathrm{B}$ neutrino flux
and the survival probability
$ P_{\nu_e\to\nu_e}(E) $,
that we have discussed here,
will allow to determine
the parameters
that characterize neutrino properties
directly from the experimental data.
The ratios $\RNC$ and $\RES$
which give the relative contribution
of $\nu_\mu$ and/or $\nu_\tau$
to the rate of NC and ES events, respectively,
depend only on the survival probabilities
and do not depend on the initial flux
of $^8\mathrm{B}$ $\nu_e$'s
(see \eqts{E15} and \eqtm{E33}).
In Fig.\ref{F1} and Fig.\ref{F2}
the results of the calculations
of $\RNC$ and $\RES$
in the framework of the two-flavour MSW mechanism
are presented.
In this case the survival probability depends only
on the two parameters $\Delta m^2$ and $\sinp{2}2\vtheta$.
The curves in Fig.\ref{F1} and Fig.\ref{F2}
correspond to fixed values of $\RNC$ and $\RES$,
respectively.
The two regions
of values of
$\Delta m^2$ and $\sinp{2}2\vtheta$
which were obtained
from the analysis of the existing data [\ref{GAL92B}]
(under the assumption that the fluxes of initial solar neutrinos
are given by the SSM)
are also plotted.

\vspace{1in}
{\Large\bf Acknowledgement}

We would like to express our gratitude to
V. de Alfaro, W. Grimus and A. Suzuki
for useful discussions
and to thank J. Bahcall
for useful communications about the SSM.

\vspace{1in}

{\Large\bf References}

\begin{list}{[\therefs]}{\usecounter{refs}}

\item\label{DAV90}
R. Davis et al.,
Proceedings of the 21$^{\mathrm{th}}$
International Cosmic Ray Conference,
Adelaide, Australia, 1990.

\item\label{KII}
K. S. Hirata et al.,
Phys. Rev. Lett. 65 (1990) 1297;
Phys. Rev. D 44 (1991) 2241.

\item\label{SAGE}
A.I. Abazov et al.,
Phys. Rev. Lett. 67 (1991) 3332;
V.N. Gavrin et al.,
presented at IV Rencontres de Blois, June 1992.
                   
\item\label{GAL92A}
GALLEX Collaboration,
Phys. Lett. B 285 (1992) 376.

\item\label{BAH}
J.N. Bahcall and R. Ulrich,
Rev. Mod. Phys. 60 (1988) 297;
J.N. Bahcall,
Neutrino Physics and Astrophysics,
Cambridge University Press, 1989;
J.N. Bahcall and M.H. Pinosseault,
Rev. Mod. Phys. 64 (1992) 885.
                   
\item\label{TUR88}
S. Turck-Chi\`eze, S. Cahen, M. Cass\'e and C. Doom,
Astrophys. J. 335 (1988) 415.

\item\label{PON58}
B. Pontecorvo,
J. Exptl. Theoret. Phys. 33 (1957) 549
[Sov. Phys. JETP 6 (1958) 429];
J. Exptl. Theoret. Phys. 34 (1958) 247
[Sov. Phys. JETP 7 (1958) 172].

\item\label{MSW}
L. Wolfenstein,
Phys. Rev. D 17 (1978) 2369;
S. P. Mikheyev and A. Y. Smirnov,
Il Nuovo Cimento C 9 (1986) 17.
                    
\item\label{GAL92B}
GALLEX Collaboration,
Phys. Lett. B 285 (1992) 390.

\item\label{SNO}
The Sudbury Neutrino Observatory Collaboration,
Phys. Lett. B 194 (1987) 321;
H.H. Chen,
Nucl. Instr. Meth. A 264 (1988) 48.

\item\label{SK}
A. Suzuki,
Proc. of the Workshop on Elementary Particle Picture of the Universe,
KEK, Japan, 1987;
Y. Totsuka, ICRR-report-227-90-20 (1990).

\item\label{BOREX}
BOREXINO Collaboration,
BOREXINO at Gran Sasso:
Proposal for a real time detector for low energy solar neutrinos,
August 1991.

\item\label{ICARUS}
ICARUS Collaboration,
ICARUS I: An Optimized Real Time Detector of Solar Neutrinos,
LNF-89/005(R) (1989).

\item\label{HELLAZ}
G. Bonvicini et al.,
Proceedings of the 5$^{\mathrm{th}}$ International Workshop
on Neutrino Telescopes,
Venezia, March 1993.
                  
\item\label{BIL}
For example,
see the reviews:
S.M. Bilenky and B. Pontecorvo,
Phys. Rep. 41 (1978) 225;
S.M. Bilenky and S.T. Petcov,
Rev. Mod. Phys. 59 (1987) 671.

\item\label{VVO}
A. Cisneros,
Astrophys. Space. Sci. 10 (1970) 87;
M.B. Voloshin, M.I. Vysotsky and L.B. Okun,
Zh. Eksp. Teor. Fiz. 91 (1986) 754
[ Sov. Phys. JETP 64 (1986) 446].
                          
\item\label{LM88}
C.S. Lim and W.J. Marciano,
Phys. Rev. D 37 (1988) 1368.
                       
\item\label{AKH88}
E.K. Akhmedov,
Phys. Lett. B 213 (1988) 64.

\item\label{SNO92}
G.T. Ewan,
Proceedings of the 4$^{\mathrm{th}}$ International Workshop
on Neutrino Telescopes,
Venezia, March 1992.

\item\label{CC}
F.J. Kelly and H. Uberall,
Phys. Rev. Lett. 16 (1966) 145;
S.D. Ellis and J.N. Bahcall,
Nucl. Phys. A 114 (1968) 636.

\item\label{BAH86}
J.N. Bahcall and B.R. Holstein,
Phys. Rev. C 33 (1986) 2121.

\item\label{BAH91}
J.N. Bahcall,
Phys. Rev. D 44 (1991) 1644.

\item\label{PAL92}
P.B. Pal,
Int. J. Mod. Phys. A 7 (1992) 5387.

\item\label{NC}
H.H. Chen,
Phys. Rev. Lett. 55 (1985) 1534;
R.S. Raghavan, S. Pakvasa and B.A. Brown,
Phys. Rev. Lett. 57 (1986) 1801;
S. Weinberg,
Int J. Mod. Phys. A 2 (1987) 301.

\item\label{BAH88}
J.N. Bahcall, K. Kubodera and S. Nozawa,
Phys. Rev. D 38 (1988) 1030.

\item\label{AKH93}
E.K. Akhmedov, S.T Petcov and A.Yu. Smirnov,
SISSA 9/93/EP.

\end{list}

\vspace{1in}

{\Large\bf Figure Captions}

\begin{list}{Fig.\thefigs}{\usecounter{figs}}

\item\label{F1}
The ratio $\RNC$,
which gives the relative contribution
of $\nu_\mu$ and/or $\nu_\tau$
to the rate of NC events.
The curves correspond
to the values of $\RNC$
equal to 0.1, 0.2, ..., 0.9.
The two regions
of values of
$\Delta m^2$ and $\sinp{2}2\vtheta$
which were obtained from the analysis of the existing data
[\ref{GAL92B}]
are also plotted.

\item\label{F2}
The ratio $\RES$,
which gives the relative contribution
of $\nu_\mu$ and/or $\nu_\tau$
to the rate of ES events.
The curves correspond
to the values of $\RES$
equal to 0.1, 0.2, ..., 0.9.
The two regions
of values of
$\Delta m^2$ and $\sinp{2}2\vtheta$
which were obtained from the analysis of the existing data
[\ref{GAL92B}]
are also plotted.

\end{list}

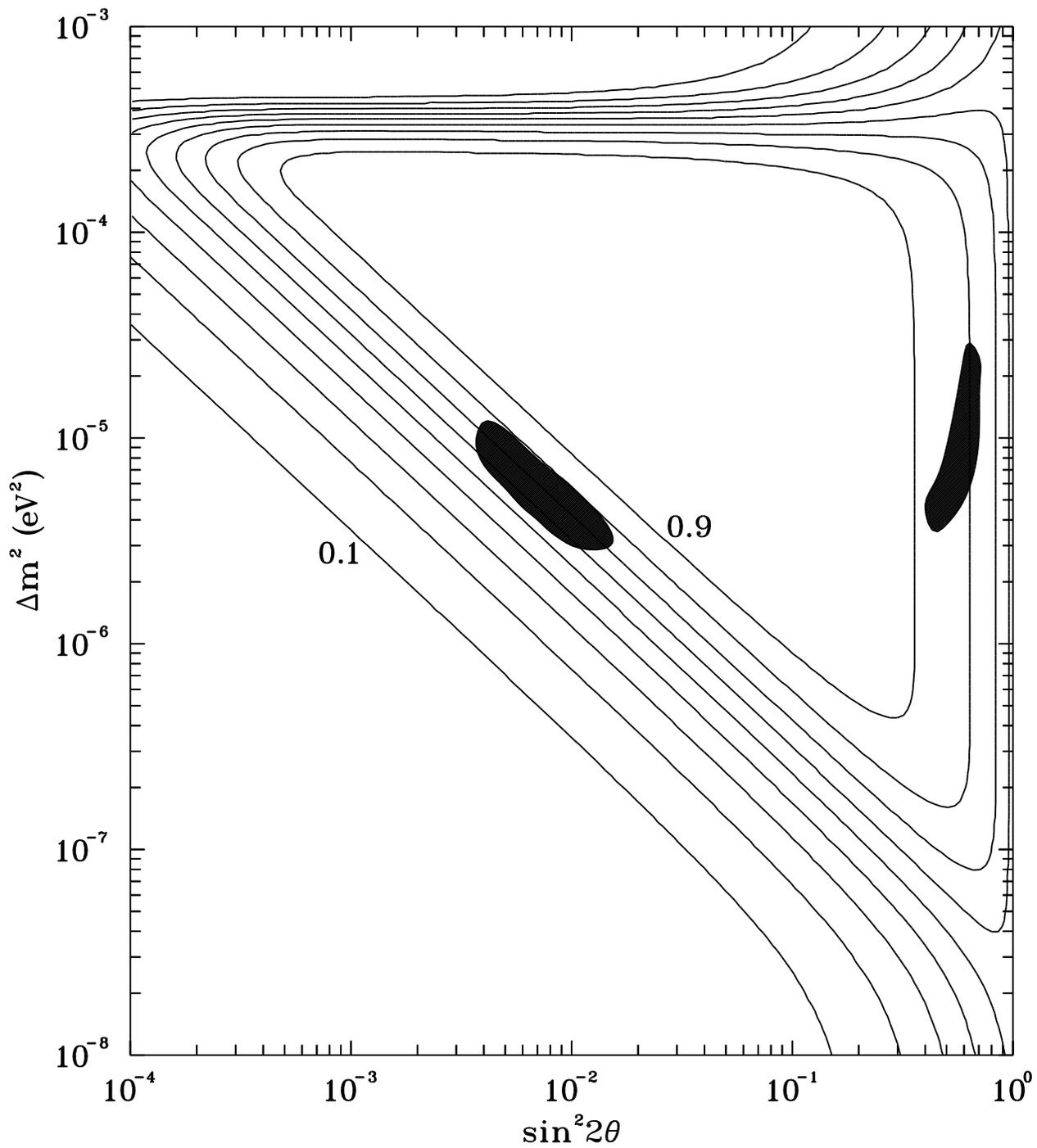
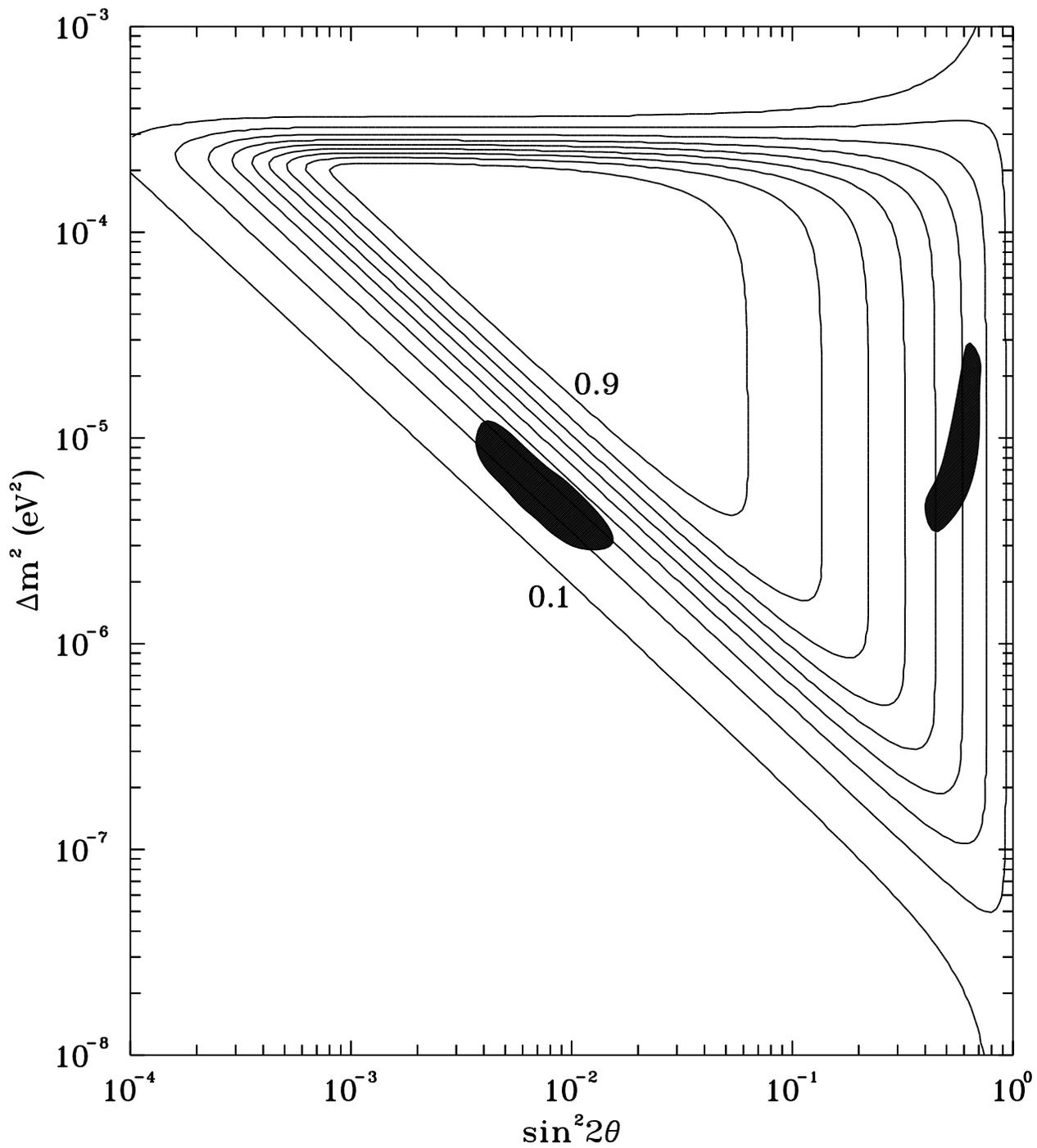
\end{document}